\documentclass[prd,twocolumn,showpacs]{revtex4}   
\usepackage{amsmath}    
\usepackage{amsfonts}  
\usepackage{amssymb}
\usepackage{graphicx}
\newcommand{\be}{\begin{equation}}
\newcommand{\ee}{\end{equation}}
\newcommand{\bea}{\begin{eqnarray}}
\newcommand{\eea}{\end{eqnarray}}
\newcommand{\nn}{\nonumber}

\begin{document}

\title{Fermion Condensate and Vacuum Current Density Induced by Homogeneous and Inhomogeneous Magnetic Fields in (2+1)-Dimensions}

\author{Alfredo Raya}
\email{raya@ifm.umich.mx}
\affiliation{Instituto de F\'{\i}sica y Matem\'aticas, Universidad
  Michoacana de San Nicol\'as de Hidalgo. Edificio C-3, Ciudad Universitaria,
  Morelia, Michoac\'an, 58040, M\'exico.}
\author{Edward Reyes}
\email{edward.reyes@nucleares.unam.mx}
\affiliation{Instituto de Ciencias Nucleares, Universidad Nacional
  Aut\'onoma de M\'exico.  Circuito Exterior s/n, Ciudad
  Universitaria, M\'exico, D.F., 04510, M\'exico.}
\altaffiliation{Also at: Instituto de F\'{\i}sica y Matem\'aticas, Universidad
  Michoacana de San Nicol\'as de Hidalgo. }

\date{\today}

\begin{abstract}

We calculate the condensate and the vacuum current density induced by  external static magnetic fields in (2+1)-dimensions. At the perturbative level, we consider an exponentially decaying magnetic field along one cartesian coordinate. Non-perturbatively, we obtain the fermion propagator in the presence of a uniform magnetic field  by solving the Schwinger-Dyson equation in the rainbow-ladder approximation.  In the large flux limit, we observe that both these quantities, either perturbative (inhomogeneous) and non-perturbative (homogeneous), are proportional to the external field, in agreement with early expectations.
\end{abstract}
\maketitle

\section{Introduction}

For massless fermion theories, several aspects of the {\em magnetic catalysis} of dynamical chiral symmetry breaking, {\em i.e.}  the formation of a fermion condensate by effects of a uniform magnetic field,  have been a subject of intense scrutiny over the past two decades~\cite{mc1}. More recently, the same effect was shown to generate an anomalous magnetic moment through a dynamically generated Bohr magneton inversely proportional to the dynamical mass~\cite{cubanos}. In this connection, aside from their role as toy models in particle physics, theories of fermions in (2+1)-dimensions have captured the interest of the community because the potential applications in condensed matter systems, for which the low energy dynamics can be described in terms of planar fermions (see, for example~\cite{Sharapov} and references therein), including graphene in the massless version~\cite{graphene}. 
These theories exhibit unique features that make them interesting on their own. For example, the ground state of an odd number of fermions in a uniform magnetic field exhibits a finite value of a fermion condensate~\cite{gusynin} and a parity non-invariant current~\cite{Khalilov,Redlich}. Parity breaking and gauge noninvariance are intimately connected for such a system~\cite{Redlich}. Gauge invariance can be restored at the expense of introducing a parity non-invariant Chern-Simons term in the effective action for fermions, or, equivalently, in the action of the gauge field. In any case, parity is explicitly broken. The formation of the condensate by effects of a magnetic field is important, for example, for planar doped antiferromagnets. Such systems are relevant to the physics of high-$T_c$ superconductors~\cite{mavro,Farakos} in the surface region of these materials, where, due to the Meissner effect, the external magnetic field can penetrate the sample.  In Ref.~\cite{Farakos}, the dynamical mass gap generated by an intense external uniform magnetic field was studied in the reducible formulation of parity-invariant quantum electrodynamics in the plane, QED3, solving the Schwinger-Dyson equation in the rainbow-ladder truncation and constant-mass approximation, in a dimensionally-reduced variant of the well know expression of magnetic catalysis in (3+1)-dimensions. In that work, the proper-time representation~\cite{schwinger} of the fermion propagator was used. The dynamical formation of this gap was found to be connected to the enhancement of the superconducting gap in the strong-$U$ Hubbard model of Ref.~\cite{mavro}. It is important to point out that the magnetic field in the surface region of these materials is not uniformly distributed in space. Therefore, it becomes important to consider spatial anisotropies in the formation of condensates. In this connection, through the second quantized solutions to the parity-invariant Dirac equation in a reducible representation of the $\gamma^\mu$-matrices, it was shown~\cite{Dunne} that for non-uniform static magnetic fields of intense flux, the inhomogeneous condensate follows the spatial profile of the applied field, in a local version of the Aharonov-Casher integrated relation~\cite{AC}, which is essentially the Landau's degeneracy-flux relation~\cite{landau}. Such a relation is  connected with mathematical index theorems~\cite{index}. The inhomogeneous fields considered in Ref.~\cite{Dunne} include fields which vary along the radial or a cartesian coordinate. The contribution to the condensate in the massless limit was found to come only from the threshold states, for which the mass equals in magnitude its energy. 
  
  In this work, we are interested in obtaining the fermion condensate and vacuum current density induced by homogeneous and inhomogeneous magnetic fields, but our approach is different. We derive these quantities directly from the fermion propagator. For this purpose, we use the so-called Ritus eigenfunctions method~\cite{Ritus}, which provides an elegant and powerful method to diagonalize the fermion Green's function in momentum space in the presence of external fields. The method was originally developed for spin-1/2 fermions, and was later extended to the case of charged bosons~\cite{boson}. In both these cases, uniform fields were considered. Here, we present a generalization of  the Ritus approach to the case of inhomogeneous fields in (2+1)-dimensions~\cite{murguia}. We assume the field aligned perpendicularly to the plane of motion of fermions. The diagonalization of the Green's function in momentum space for uniform and non-uniform magnetic fields might also be useful for the theoretical studies derived from the measurement of half integer quantum Hall effect and zero-energy Landau level in graphene at room temperature~\cite{novo}. Moreover, the method allows to write down general expressions for the condensate and the induced current density presented in Eqs.~(\ref{condensado2})-(\ref{corr2}). We consider two special field configurations. First we implement Ritus method to express the tree-level fermion propagator in the presence of an exponentially damped magnetic field along one spatial dimension. Such a field occurs, to a good approximation, inside the London penetration depth of a type-I superconductor if a semiconductor heterostructure -with narrow quantum well- is introduced perpendicularly to the planar surface of the superconductor into a narrow slit, and a homogeneous magnetic field is applied parallel to the surface of the superconductor~\cite{solsexp1}. As a second example, we obtain the non-perturbative fermion propagator in QED$_3$ with the same assumptions that Ref.~\cite{Farakos}, but expanding the propagator in the Ritus eigenfunctions. From this propagator we derive the non-perturbative condensate and induced charge density.
   This article is organized as follows: in Sect.~\ref{fermions}, we describe the Lagrangian for Dirac fermions in (2+1)-dimensions and in the presence of external fields, with emphasis in the symmetry properties of the mass terms. Section~\ref{propagator} is devoted to the fermion propagator in the presence of an inhomogeneous magnetic field which is exponentially damped along one spatial direction~\cite{murguia} within the Ritus formalism~\cite{Ritus}. The perturbative condensate and induced vacuum current density for this  field are discussed. The non-perturbative propagator in a uniform magnetic field and the corresponding homogeneous condensate and vacuum charge density are discussed in Sect.~\ref{uniform}. Final remarks are presented in Sect.~\ref{Final}.

\section{Planar Fermions}
\label{fermions}

We start from the Dirac Lagrangian in external electromagnetic fields
\be
{\cal L}_D=\bar\psi(\gamma\cdot\Pi-m)\psi, \label{Diraclag}
\ee
where $\Pi_\mu=i\partial_\mu+eA_\mu$. For a more detailed presentation of the symmetries of this Lagrangian, see, for instance, Ref.~\cite{planarfermions}.
In (2+1)-dimensions, only three Dirac matrices are required to fulfill the Clifford algebra $\{\gamma^\mu,\gamma^\nu\}=2g^{\mu\nu}$. The lowest dimensional representation of these matrices is $2\times 2$ and hence we can choose them to be proportional to the Pauli matrices as 
\be
\gamma^{0} = \sigma_3, \quad \gamma^{1} =
i\sigma_1, \quad \gamma^{2} = i\sigma_2 . \label{primera} 
\ee
We call this representation ${\mathcal A}$. In this representation, fermions posses only one spin orientation. Additionally, there exists a second inequivalent representation, labeled ${\mathcal B}$, which can be chosen as follows
\be
\tilde{\gamma}^{0} = \sigma_3, \quad \tilde{\gamma}^{1} = i\sigma_1, \quad \tilde{\gamma}^{2} =
-i\sigma_2, \label{segunda}
\ee
in which fermions have the opposite  spin orientation alone. In graphene, representations ${\mathcal A}$ and ${\mathcal B}$ are required to describe two different species of massless fermions in each triangular sub-lattice of the honeycomb lattice~\cite{Semenoff}. Chiral symmetry cannot be defined for either ${\mathcal A}$ or ${\mathcal B}$, because there is no $2\times 2$ matrix analogous to `$\gamma_5$'. Moreover, the mass term $m\bar\psi\psi$ in the Lagrangian is non-invariant under the parity transformation $x_1\to-x_1$ and $A_1(x_1,x_2)\to -A_1(-x_1,x_2)$ for these representations.

The two species with their respective spin orientations can be conveniently combined into a 4-component spinor with a $4\times 4$ representation of the Dirac matrices, say,
\begin{eqnarray}
\gamma^0= \left(\begin{array}{cc} \sigma_3 & 0 \\ 0 & -\sigma_3
\end{array}\right), \qquad {\gamma^k}=\left(\begin{array}{cc}i \sigma_k & 0  \\ 0  & -i \sigma_k
\end{array}\right),\label{reducible}
\end{eqnarray}
 for $k=1,2$, which we label ${\mathcal C}$, and
\begin{eqnarray}
\gamma^3= i\left(\begin{array}{cc} 0 & \mathbb{I} \\ \mathbb{I} & 0
\end{array}\right), \qquad \gamma^{5}= i\gamma^0\gamma^1\gamma^2\gamma^3=i \left(\begin{array}{cc} \phantom{-}0 & \mathbb{I}  \\ -\mathbb{I}  & 0
\end{array}\right).
\end{eqnarray}
Here ${\mathbb I}$ is the identity matrix (we use the same symbol in any dimensionality). In this reducible representation, $m\bar\psi\psi$ is parity invariant. Furthermore, in the massless limit, the Lagrangian~(\ref{Diraclag}) possesses a global $U(2)$ flavor symmetry, with generators ${\mathbb I}$, $\gamma^3$, $\gamma^5$ and $[\gamma^3,\gamma^5]$, corresponding to the interchange of the two irreducible species.  The ordinary mass term breaks this symmetry. However, there exists a second mass term of the form $m_o\bar\psi\tau\psi$, with  $\tau= \lbrack \gamma^3,\gamma^5 \rbrack/2={\rm diag}(\mathbb{I},-\mathbb{I})$, which in condensed matter literature is often referred to as Haldane mass term~\cite{Haldane}. This mass term is invariant under  flavor symmetry, but breaks parity. In this case, the parity non-invariant Dirac Lagrangian takes the form
\be
{\cal L}_D=\bar\psi(\gamma\cdot\Pi-m-m_o\tau)\psi. \label{Diraclagred}
\ee
In order to explicitly separate the physical fermion content of this Lagrangian, we introduce the chiral-like projectors
$ \chi_\pm = (\mathbb {I} \pm \tau)/2 $ and define the ``right-handed'' $\psi_+$ and ``left-handed'' $\psi_-$ fields as $\psi_\pm=\chi_\pm \psi$. Then, the Lagrangian acquires the form
\be
{\cal L}_D=\bar\psi_+(\gamma\cdot\Pi-m_+)\psi_+ + \bar\psi_-(\gamma\cdot\Pi-m_-)\psi_-, \label{Diraclagredsep}
\ee
where $ m_\pm = m\pm  m_o$. Thus, in this form, the Lagrangian is neatly seen to describe two different fermion species, and the effect of the parity-violating mass term is to remove the mass degeneracy between them. 
Below we obtain the fermion propagator for Lagrangians~(\ref{Diraclag}) and~(\ref{Diraclagredsep}) in external magnetic fields.

\section{Propagator in Inhomogeneous Magnetic Fields}
\label{propagator}

In this section we obtain the fermion propagator in an inhomogeneous magnetic field perpendicular to the plane of motion of the fermions in (2+1)-dimensions within the Ritus formalism~\cite{Ritus}. Working in a Landau-like gauge, we choose $A_\mu =(0,0,W(x))$ such that the profile of the field, which we consider varying along the $x$-axis, is $B(x)=W'(x)=\partial_xW(x)$. Details of this derivation can be found in Ref.~\cite{murguia}. The Green's function for the Dirac equation, $S(z,z')$, satisfies
\be
(\gamma\cdot\Pi-m)S(z,z')=\delta^{(3)}(z-z'), \label{green}
\ee
with $z=(t,x,y)$. Since $ S(z,z') $ commutes with $ (\gamma \cdot \Pi)^2 $, we expand it on the basis of its eigenfunctions,  
\be
(\gamma \cdot \Pi)^2  \mathbb{E}_p(z)= p^2 \mathbb{E}_p(z).
\ee
Without loss of generality, we work only with the irreducible representation ${\mathcal A}$, and specify how results are modified in other representations.  
In this case, $(\gamma \cdot \Pi)^2=\Pi^2+e\sigma_3W'(x)$, and thus
the Ritus eigenfunctions become
\be
{\mathbb E}_p^{({\mathcal A})}(z)=
\begin{pmatrix} E_{p,+1}(z) & 0 \\ 0 & E_{p,-1}(z)\end{pmatrix}.
\label{ritus}
\ee
With the aid of the property $\gamma^\mu\gamma^\nu=g^{\mu\nu}-i\epsilon^{\mu\nu\lambda}\gamma_\lambda$, these can be expressed in the more convenient form
\be
{\mathbb E}_p^{({\mathcal A})}(z)= E_{p,+1}(z) \Delta(+)+ E_{p,-1}(z)\Delta(-)\;,
\ee
with 
\be
\Delta(\pm)= \frac{{\mathbb I}\pm i\gamma^1\gamma^2}{2}\;,
\ee
being the spin projectors~\cite{cubanos}. In the above expressions, the subscript $ p=(p_0,p_2, k)$ denotes the eigenvalues of the operators
$i\partial_t$, $-i\partial_y$, and ${\cal H}=-(\gamma\cdot \Pi)^2 + \Pi_0^2$,
respectively, $ \sigma = \pm 1 $ are the eigenvalues of $\sigma_3 $ and
\be
E_{p,\sigma}(z)=N_\sigma e^{-i(p_ot-p_2y)}F_{k,p_2}^\sigma(x), \label{an2}
\ee
with $N_\sigma$ a normalization constant. $F_{k,p_2}^\sigma(x)$ satisfies
\be
\left[\partial_x^2-(-p_2+eW(x))^2+e\sigma W'(x)+k\right] F_{k,p_2}^\sigma(x)= 0,\label{pauli}
\ee
which is the equation of the Pauli Hamiltonian with constrained vector potential, mass $m=1/2$ and gyromagnetic factor $g=2$. This Hamiltonian turns out to be supersymmetric in the sense of SUSY-quantum mechanics~\cite{susyqm}. $F_{k,p_2}^{+1}(x)$ and $F_{k,p_2}^{-1}(x)$ are the solutions of the respective SUSY-partner potentials
\be
V_\pm(x)=(-p_2+eW(x))^2\pm eW'(x).\label{susypot}
\ee 
Regrettably, the solution to Eq.~(\ref{pauli}) for arbitrary $W(x)$ is unknown. For certain inhomogeneous fields which are translationally invariant along one spatial direction, the solutions can be expressed in terms of an orthogonal system of functions~\cite{susyqm}. However, the following important property of the Ritus eigenfunctions,
\be
(\gamma  \cdot \Pi) \mathbb{E}_p(z)= \mathbb{E}_p(z) (\gamma \cdot \bar{p}),  \label{vector}
\ee
where the three-momentum vector  $\bar{p}$ satisfies $\bar{p}^2=p^2=p_0^2-k$ with  $ \bar{p}_\mu = (p_0, 0, \sqrt {k})$ is valid in the general case~\cite{murguia}. The explicit matrix form of Eq.~(\ref{vector}) is 
\begin{eqnarray}
\left(\begin{array}{cc} i\partial_t E_{p,+1}(z) & D_-E_{p,-1}(z) \\ D_+ E_{p,+1}(z) & -i\partial_t E_{p,-1}(z)
\end{array}\right)  \nn\\
&& \hspace{-3.5cm}
=\left(\begin{array}{cc} p_0 E_{p,+1}(z) & -\sqrt{k} E_{p,+1}(z) \\ \sqrt{k}E_{p,-1}(z) & -p_0 E_{p,-1}(z)
\end{array}\right)\;,
\end{eqnarray}
with $D_\pm=-\partial_x\pm(-i\partial_y-eW(x))$. The relations following from diagonal components of this matrix equation are directly inferred from the properties of the $E_{p,\sigma}(z)$ functions, while the off-diagonal components can be cast in the form of the `kinetic-balance' system of equations
\bea
D_- E_{p,-1}(z) &=&-\sqrt{k} E_{p,+1}(z),\nn\\
D_+ E_{p,+1}(z) &=& \sqrt{k} E_{p,-1}(z). \label{desenso}
\eea
Thus the $E_{p,\sigma}(z)$ functions which satisfy these expressions also satisfy the identity~(\ref{vector}). Its explicit form, however, will depend on the field under consideration.

\subsection{Exponentially Decaying Magnetic Field}

In order to proceed further, we concentrate the discussion to the case of an exponentially decaying magnetic field $B(x)=Be^{-\hat\alpha x}$. Solutions to the Dirac equation in this field  has been studied in Ref.~\cite{solsexp1,solsexp2,solsexp3}. 
Such a field occurs inside the penetration depth of a type-I superconductor when a semiconductor heterostructure with narrow quantum well is introduced perpendicularly to the planar surface of the superconductor, and a homogeneous magnetic field is applied parallel to this  surface~\cite{solsexp1}.
An exponentially decaying field can be described by $W(x) =-(B/\hat\alpha) [\exp \{- \hat\alpha x \}$-1]. This choice allows to directly recover the uniform magnetic field case of Sect.~\ref{uniform} by setting $\hat\alpha=0$. Defining the dimensionless variables~\cite{solsexp1,solsexp2}
\be 
\xi \equiv \frac{eB}{{\hat\alpha}^2} e^{-\hat\alpha x}=\frac{1}{[\hat\alpha \ell(x)]^2}\;,
\ee
 and 
\be
 s=\frac{|\hat{p}_2|}{\hat\alpha}\equiv\frac{eB}{{\hat\alpha}^2} e^{-\hat\alpha x_0} =\frac{1}{[\hat\alpha \ell(x_0)]^2} \;,
\ee
where $\ell(x)$ is the magnetic length, $\ell(x_0)$ its local counterpart~\cite{solsexp1,solsexp2} and $\hat{p}_2=p_2+eB/\hat\alpha$, the Pauli equation~(\ref{pauli}) takes the form
\be
\left[ \frac{\partial^2}{\partial \xi^2} + \frac{1}{\xi} \frac{\partial}{\partial \xi} - \frac{s^2-\varepsilon^2}{\xi^2}+ \frac{(2s+\sigma)}{\xi} -1\right] F_{k,p_2}^{\sigma}(\xi)=0 \;.\label{pauliexp}
\ee
Here, $\varepsilon^2=k/\hat\alpha^2$. The behavior at small and large $\xi$ of this equation suggests $F_{k,p_2}^{\sigma}(\xi)\sim\xi^\beta e^{-\xi} \omega^\sigma(\xi)$, with $\beta^2=s^2-\varepsilon^2$. Inserting this ansatz, Eq.~(\ref{pauliexp}) becomes
\be
\left[\xi\frac{d^2}{d\varrho^2}+(\lambda-\varrho)\frac{d}{d\varrho}-\eta_\sigma\right]\omega^\sigma(\varrho)=0\;,\label{confluent}
\ee
with $\lambda=2\beta+1$ and $\eta_\sigma=-s+\beta+1/2-\sigma/2$ and $\varrho=2\xi$.
The general solution to this equation is $\omega^\sigma=c_{(1)} \  _1F_1[\eta_\sigma,\lambda,\varrho]+c_{(2)} U[\eta_\sigma,\lambda,\varrho]$, where $_1F_1[a,b,x]$ and $U[a,b,x]$ are, respectively, confluent hypergeometric functions of the first and second kind. The irregular behavior of $U[a,b,x\to 0]$ enforces $c_{(2)}=0$. Moreover, to preserve the asymptotic behavior, it is required that $\eta_\sigma$ is a negative integer, $-|\eta_\sigma|=0,1,2,\ldots$ This implies that 
\be
k_\sigma=\hat\alpha^2\left\{s^2-\left[s-\left(|\eta_\sigma|+\frac{1}{2}-\frac{\sigma}{2}\right)\right]^2\right\}\;.
\ee
The energy eigenvalues for a particle on-shell, $p^2=m^2$, are conveniently written as
\be
k_n=\hat{p}_2^2-(\hat{p}_2-n\hat\alpha)^2\label{energias}.
\ee
Notice that $k_n$ explicitly depends on the momentum $\hat{p}_2$, unlike the case of the uniform magnetic field case, even in (3+1)-dimensions. For such a particle, the energy $p_0^2=k_n+m^2$ cannot be tachyonic. This fact restricts $\beta=s-n>0$, which can be achieved so long as  $\hat{p}_2>\hat\alpha n$. We observe that for $k_{+1}$, we can write $-|\eta_{+1}|=n-1$, and for $k_{-1}$, $-|\eta_{-1}|=n$. Regarding the solutions $\omega^\sigma(\varrho)$, for $n=0$, $\omega^{+1}_0(\varrho)$  does not exist, whereas $\omega^{-1}_0(\varrho)$ is nondegenerate. With all the above, defining $\omega_{n-1}^\sigma(\varrho)=0$, the solutions to Eq.~(\ref{confluent}) acquire the form 
\be
\omega^\sigma_n(\varrho)=\ _1F_1\left[-\left(n-\frac{1}{2}-\frac{\sigma}{2}\right),2\beta+1,\varrho\right]\;.
\ee
From the identity
\be
L_m^\kappa(x)=\frac{(\kappa+1)_m}{m!} \ _1F_1[-m,\kappa+1,x]\;,
\ee
where $L_m^{\kappa}(x)$ are the associated Laguerre polynomials and $(\kappa)_m=\Gamma(\kappa+m)/\Gamma(\kappa)$ is the Pochhammer symbol, 
solutions to Eq.~(\ref{pauliexp}) can be expressed as~\cite{solsexp3}
\be
F_{k,p_2}^{\sigma}(\varrho) = \psi_{n-\frac{1}{2}-\frac{\sigma}{2}}^{2s-2n}(\varrho),\label{soluciones}
\ee
where
\be
\psi_m^\kappa(x) = e^{-x/2} x^{\kappa/2} L^{\kappa}_m(x),
\ee
are the Laguerre functions~\cite{arfken}, which form a complete set of orthogonal functions in the interval $(0,\infty)$ and verify the orthogonality relation
\be
\int_0^\infty dx \ \psi_{m}^\kappa (x)\psi_{m'}^\kappa(x)= \frac{\Gamma(m+\kappa+1)}{n!} \ \delta_{m,m'}.
\ee
This relation is important to understand the spectrum of bound states of Eq.~(\ref{pauliexp}) below. 

\begin{center}
\begin{figure}
\includegraphics[width=0.45\textwidth]{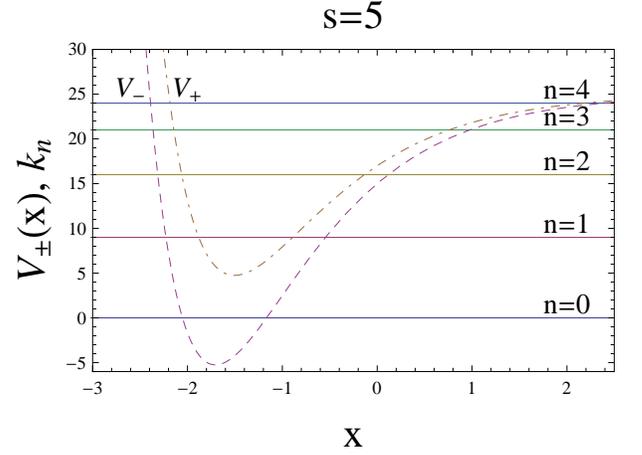}
\caption{SUSY-parter potentials~(\ref{susypot}) and energy eigenvalues for the exponentially decaying magnetic field  for $s=5$. The number of bound states grows linearly with $s$. The scale of the plot is set by $\hat\alpha=eB=1$.}
\label{fig1}
\end{figure}
\end{center}

Now, the energy eigenvalues in Eq.~(\ref{energias}) suggest the identification of the quantum number $n$ with the Landau level index. This is indeed the case, because such eigenvalues, in the limit  $\hat\alpha\to 0$,  reduce to the Landau levels for a uniform magnetic field, 
\be
k_n=w_c^2(x_0) 2n \;,
\ee
where $w_c(x_0)=1/\ell(x_0)$ is the local cyclotron frequency. Moreover, because $s$ is a real parameter, the number of normalizable solutions for a given $s$, restricts the Landau level index $n$ to take the values $n=0,1,2,\ldots,\varsigma-1$ where $\varsigma=[s]$ is the integer part of $s$. This is evident from Fig.~\ref{fig1}, where the SUSY-partner potentials~(\ref{susypot}) are shown as a function of $x$ for fixed $s=5$. The energy eigenvalues $k_n$ for the normalizable states are also displayed.

With all the above, the solutions $E_{p,\sigma}(z)$ in Eq.~(\ref{an2}) acquire the explicit forms 
\bea
E_{p,+1}(z) &=& \frac{\hat\alpha}{2\pi}\left( \frac{2  n! (s-n)}{\Gamma (2s-n+1)} \right)^{1/2} e^{-ip_0 t + i p_2 y} \nn\\
&&\times e^{-\varrho/2} \varrho^{(s-n)}L_{n}^{2(s-n)} (\varrho) , \nn\\
E_{p,-1}(z) &=& \frac{\hat\alpha}{2\pi}\left( \frac{2  (n-1)! (s-n)}{\Gamma (2s-n)} \right)^{1/2} e^{-ip_0 t + i p_2 y}\nn\\ &&\times e^{-\varrho/2} \varrho^{(s-n)}L_{n-1}^{2(s-n)} (\varrho). \label{exponenciales}
\eea
Inserting these expressions into Eq.~(\ref{ritus}), we obtain the Ritus eigenfunctions for an exponentially decaying static magnetic field.
It is straightforward to check that these  verify the orthogonality and closure relations
\bea
\nn
\int d^3z \ \overline {\mathbb {E}}_{p'}(z) \mathbb {E}_p(z) = \hat{\delta}^{(3)}(p-p')\Pi(n), \\
{\int\!\!\!\!\!\!\!\!\sum} d^3p \ \mathbb {E}_{p}(z) \overline {\mathbb {E}}_p (z') = \delta^{(3)}(z-z'), \label{ortonormalidad}
\eea
where $ \overline {\mathbb{E}}_p = \gamma^0 \mathbb {E}_p^\dagger \gamma^0 $,
\be
\hat{\delta}^{(3)}(p-p')=\delta_{n,n'}\delta_{s,s'}\delta(p_0-p_0'), 
\ee
and the projector~\cite{cubanos,wang}
\be
\Pi(n)=\Delta(+)\delta_{n,0} +{\mathbb I}(1-\delta_{n,0}),
\ee
determines only one spin projection in the lowest Landau level (LLL). The symbol $\int \!\!\!\!\!\!\sum d^3p$ indicates that the integration might represent a sum, depending upon the continuous or discrete nature of the components of  $p=(p_0,p_2,k)$.
In our example,
\be
{\int \!\!\!\!\!\!\!\! \sum} d^3p = -\hat\alpha \int dp_0 \sum_{\varsigma} \sum_{n}\;.
\ee 
Equations~(\ref{ortonormalidad})
follow directly from the complete and orthogonal character of the solutions~(\ref{exponenciales}).

\subsection{Condensate and Induced Electric Current}

Physically, Ritus eigenfunctions $ \mathbb {E}_p(z)$ correspond to the asymptotic states of electrons with momentum $ \bar {p} $ in the external field. Therefore, we can use these functions to expand $S(z,z')$ in momentum space in the same way plane waves are used to define the Fourier transform,  
\be
S(z,z')= {\int\!\!\!\!\!\!\!\!\sum} d^3p \ d^3p' \ {\mathbb E}_p(z) S(p,p') \overline{\mathbb E}_{p'}(z')\;. \label{integ}
\ee
Inserting this Green's function in Eq.~(\ref{green}), using the properties~(\ref{vector}) and~(\ref{ortonormalidad}), the propagator in momentum space takes the form 
\be
S(p,p')=\hat{\delta}^{(3)}(p-p')\Pi(n)\tilde{S}(\overline{p}),
\ee
where
\be
\tilde{S}(\overline{p})=\frac{1}{\gamma\cdot \overline{p}-m}, \label{propagador_libre}
\ee
with $\bar{p}=(p_0,0,\sqrt{\hat{p}_2^2-(\hat{p}_2-n\hat\alpha)^2})$.  This simple form allows a direct calculation of physical observables, like
the fermion condensate 
\bea
\langle\bar\psi\psi\rangle =  Tr \{i S(z,z)\}, \label{condensado}
\eea
and the induced vacuum current density 
\bea
j^\mu =-ieTr \{ \gamma^\mu S(z,z) \}. \label{corriente}
 \eea
These acquire the general form
\bea
\langle \bar {\psi} \psi \rangle_{\mathcal A} 
&=& i \ {\int \!\!\!\!\!\!\!\! \sum}d^3p  \frac {m}{\overline{p}^2 - m ^2} \nn\\
&&\times \left [  \vert  E_{p,+1}(z) \vert^2  +  \vert  E_{p,-1}(z) \vert ^2\right], \label{condensado2} \\
j^0_{\mathcal A} 
&=& -ie \ {\int\!\!\!\!\!\!\!\!\sum}d^3 p \frac {m}{\overline{p}^2  - m^2}\nn\\
&&\times  \left[\vert E_{p,+1}(z)\vert^2 - \vert E_{p,-1}(z)\vert^2\right], \label{corriente2} \\
j^k_{\mathcal A}&=&0,~~~~~k=1,2.\label{corr2}
\eea
We want to emphasize that these expressions are valid for any profile of the magnetic field so long as we know the solutions to the Pauli equation (\ref{pauli}), from which we can build the functions $E_{p, \sigma}(z) $ of Eq.~(\ref{an2}).

Inserting the explicit solutions, the condensate and charge density are
\bea
\langle \bar {\psi} \psi \rangle_{\mathcal A}&=&\frac{m\hat\alpha^2}{2\pi } \nn\\
&&\times\left \{ \sum_{\varsigma=0}^{\infty}  \frac{1}{\vert m\vert}  \left ( \frac{\varsigma}{\Gamma (2\varsigma+1)}\right )e^{-\varrho}\varrho^{2\varsigma} \lbrack L_0^{2\varsigma}(\varrho)\rbrack ^2  \right.\nn\\
\nn
&&+ \sum_{n=1}^\infty \sum_{\varsigma=n+1}^{\infty} \frac{e^{-\varrho}\varrho^{2(\varsigma-n)}}{\sqrt{\hat\alpha^2(2\varsigma n-n^2)+m^2} }   \nn\\
&&  \left [ \left( \frac{n!(\varsigma-n)}{\Gamma (2\varsigma-(n-1))}   \right) \lbrack L_n^{2(\varsigma-n)}(\varrho)\rbrack ^2  \right.\nn \\
&&\left.\left. +\left( \frac{(n-1)!(\varsigma-n)}{\Gamma (2\varsigma-n)}   \right)  \lbrack L_{(n-1)}^{2(\varsigma-n)}(\varrho)\rbrack ^2     \right]  \right\},  \\
j^0_{\mathcal A}&=&-\frac{em\hat\alpha^2}{2\pi}\nn\\
&&\times\left \{ \sum_{\varsigma=0}^{\infty}   \frac{1}{\vert m \vert } \left( \frac{\varsigma}{\Gamma (2\varsigma+1)}\right ) e^{-\varrho}\varrho^{2\varsigma}   \left \lbrack L_0^{2\varsigma}(\varrho)  \right \rbrack ^2 \right.\nn\\
&&+\sum_{n=1}^{\infty}\sum_{\varsigma=n+1}^{\infty}   \frac{ e^{-\varrho}\varrho^{2(\varsigma-n)}}{\sqrt{\hat\alpha^2(2\varsigma n-n^2)+m^2} }             \nn\\
&&\left[\left( \frac{n!(\varsigma-n)}{\Gamma (2\varsigma-(n-1))}   \right)\lbrack L_n^{2(\varsigma-n)}(\varrho)\rbrack ^2    \right.\nn \\
&&-\left. \left.\left( \frac{(n-1)!(\varsigma-n)}{\Gamma (2\varsigma-n)}   \right)  \lbrack L_{(n-1)}^{2(\varsigma-n)}(\varrho)\rbrack ^2     \right]\right\}.
\eea
Here we have only integrated over normalizable states and separated explicitly the contribution from the LLL. 
Now, because $\varsigma$  takes only discrete values, then Landau levels are highly degenerated, except the LLL. Therefore, for fields of moderated flux, we need to perform and regularize the remaining sums.  However, in the case of intense field, the leading contribution comes from the LLL.  So, in this regime, as $m\to 0$,
\bea
j^0_{\mathcal A}&=&-\frac{e\hat\alpha^2}{4\pi} \varrho e^{-\varrho} \sinh (\varrho) {\rm sgn} (m), \label{carga_exponencial} \\
\langle \bar {\psi} \psi \rangle_{\mathcal A}^{n=0,\sigma=1} &=& \frac{\hat\alpha^2}{4\pi} \varrho e^{-\varrho} \sinh (\varrho) {\rm sgn} (m)\;. \label{condensado_exponencial}
\eea
In the intense flux limit, the above expressions become
\bea
j^0_{\mathcal A} &=& - \frac{e^2}{4\pi} B e^{-\hat\alpha x} {\rm sgn} (m), \\
\langle \bar {\psi} \psi \rangle^{n=0,\sigma=1}_{\mathcal A} &=&  \frac{e}{4\pi} B e^{-\hat\alpha x} {\rm sgn} (m),\label{main}
\eea
and comprise the main results of this section.

Translation of these findings to representation ${\mathcal B}$  of Dirac matrices given in Eq.~(\ref{segunda}) is straightforward. Ritus eigenfunctions can be constructed as
\bea
\mathbb {E}^{({\mathcal B})}_p(z)=
\left(\begin{array}{cc}
E_{p,-1}(z) & 0 \\ 0 & E_{p, +1}(z)
\end{array} \right) \;,
\eea
and then, we find  that
$\langle \bar {\psi} \psi \rangle_{\mathcal B} =  \langle \bar {\psi} \psi \rangle_{\mathcal A}$, 
$j^0_{\mathcal B} =  - j^0_{\mathcal A}$, and $j^k_{\mathcal A} = j^k_{\mathcal B} = 0$ for $k = 1,2$, whereas for the reducible representation ${\mathcal C}$, Eq.~(\ref{reducible}),  the Ritus eigenfunctions are
\be
\mathbb{E}_p^{({\mathcal C})}(z)=\left(\begin{array}{cc}
{\mathbb E}_p^{({\mathcal A})}(z) & 0 \\ 0 & {\mathbb E}_p^{({\mathcal A})}(z)
\end{array} \right) \;.
\ee
The block-diagonal structure emphasizes the existence of two fermion species, each with a different mass. 
Thus, formally, $\langle \bar {\psi} \psi \rangle_{\mathcal C} =\langle \bar {\psi} \psi \rangle_{\mathcal A}(m_+)+\langle \bar {\psi} \psi \rangle_{\mathcal A}(m_-)$ and $j^\mu_{\mathcal C}=j^\mu_{\mathcal A}(m_+)+j^\mu_{\mathcal A}(m_-)$. At the end, we take the massless limit $m_\pm\to 0$ of these expressions.

\section{Propagator in Uniform Magnetic Fields}
\label{uniform}

In this section, we obtain the fermion propagator in a uniform magnetic field by solving the corresponding Schwinger-Dyson equation (SDE) in QED3. As a first step, we construct the Ritus eigenfunctions with the procedure outlined earlier. A uniform magnetic field can be specified by the choice  $W(x) =Bx $. Then, we simplify the Pauli equation~(\ref{pauli}) replacing $ k \to 2 \vert eB \vert k $ and making the change of variable $ \eta = \sqrt{2 \vert eB \vert} [x-p_2/(eB)] $, obtaining
\bea
\left [\frac{\partial^2}{\partial \eta^2} + k + \frac {\sigma}{2} {\rm sgn} (eB) - \frac {\eta^2}{4} \right ] F_{k,p_2}^{\sigma}(\eta)=0
\label{parabolicas}\;.
\eea
Solutions are parabolic cylinder functions $D_n(x)$ of order $n=k+\sigma {\rm sgn} (eB)/2-1/2$. The normalized $E_{p,\sigma}$ are
\bea
\nn
E_{p,+1}(z) &=& \frac{(\pi \vert e B \vert)^\frac{1}{4}}{2 \pi^{3/2}k!^\frac{1}{2}} e^{-ip_0 t + i p_2 y} D_k (\eta) \;, \\
E_{p,-1}(z) &=& \frac{(\pi \vert e B \vert)^\frac{1}{4}}{2 \pi^{3/2}(k-1)!^\frac{1}{2}} e^{-ip_0 t + i p_2 y} D_{k-1} (\eta).
\eea
From these, we can build up the Ritus eigenfunctions in ${\mathcal A}$ representation, Eq.~(\ref{ritus}). We use these functions to solve the Schwinger-Dyson equation for the fermion propagator in the rainbow-ladder approximation~\cite{Farakos}.

\begin{center}
\begin{figure}
\includegraphics[width=0.45\textwidth]{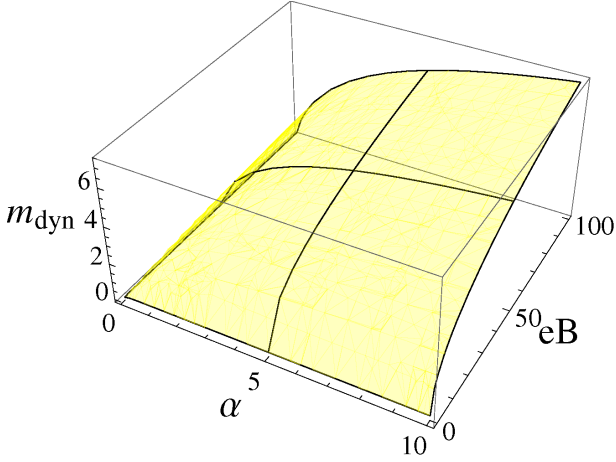}
\caption{Dynamical mass in the constant mass approximation, Eq.~(\ref{mdyn}) as a function of $\alpha$ and $eB$. }
\label{fig2}
\end{figure}
\end{center}

To this end, let us recall that the full fermion propagator verifies
\be
[\gamma\cdot\Pi-\Sigma(z,z')]G(z,z')=\delta^{(3)}(z-z'), \label{greenfull}
\ee
where $\Sigma(z,z')$ is the fermion self-energy. In Ritus formalism, the full propagator is expressed as
\be
G(z,z')= {\int\!\!\!\!\!\!\!\!\sum} d^3p \ d^3p' \ {\mathbb E}_p(z) G(p,p') \overline{\mathbb E}_{p'}(z')\;,
\ee
with 
\be
G(p,p')=\hat{\delta}^{(3)}(p-p')\Pi(n)\tilde{G}(\overline{p}),
\ee
and
\be
\tilde{G}(\overline{p})=\frac{1}{\gamma\cdot \overline{p}-\tilde{\Sigma}(\overline{p})},
\ee
where
\be
\tilde{\Sigma}(\overline{p})= \gamma\cdot \overline{p} \ Z(\overline{p})+M(\overline{p}).
\ee
On the other hand, in the rainbow-ladder approximation, the self-energy takes the form
\be
\Sigma(z,z')=-ie^2\gamma^\mu G(z,z') \gamma^\nu D_{\mu\nu}(z-z'),\label{rainbow}
\ee
where 
\bea
D_{\mu\nu}(z-z')&=&\int \frac{d^3q}{(2\pi)^3}\frac{e^{-iq\cdot (z-z')}}{q^2-i\epsilon}\nn\\
&&\times \left( g_{\mu\nu}+(\lambda-1)\frac{q_\mu q_\nu}{q^2}\right)
\eea
is the bare photon propagator and $\lambda$ is the covariant gauge fixing parameter. 
Ritus eigenfunctions allow to diagonalize the self-energy in momentum space as
\bea
\Sigma(p,p')&=&\int  d^3z d^3z' \ \overline{\mathbb {E}}_p (z) \Sigma(z,z') {\mathbb {E}}_{p'}(z') \nn\\
&=& \hat\delta^{(3)}(p-p')\Pi(n)\tilde{\Sigma}(\overline{p}).
\eea 
Combining these ingredients, Eq.~(\ref{rainbow}) becomes
\bea
\int  d^3z d^3z' \ \overline{\mathbb {E}}_p (z) \Sigma(z,z') {\mathbb {E}}_{p'}(z') &=& \nn\\
&&\hspace{-5cm}
-ie^2 \int  d^3z d^3z'  D_{\mu\nu}(z-z')\overline{\mathbb {E}}_p (z)\gamma^\mu\times\nn\\
&&\hspace{-5cm}
 \Bigg[ {\int\!\!\!\!\!\!\!\!\sum} d^3p \ d^3p' \ {\mathbb E}_p(x) G(p,p') \overline{\mathbb E}_{p'}(z')\Bigg]\gamma^\nu {\mathbb {E}}_{p'}(z')\;.
\eea
In Feynman gauge, $Z(\overline{p})=1$. Moreover, we consider the LLL approximation, and neglect the momentum dependence of the self-energy in the SDE equation, {\em i.e.,} we assume the so-called constant mass approximation, $M(0)=m_{dyn}$. In doing so, we only take into account the part of the self-energy which is proportional to the identity matrix. There is no general principle which guarantees the validity of this approximation. However, in in Ref.~\cite{Farakos}, this approximation has been established to be relieble in QED3 through the Schwinger proper-time  approach. Here we work under the same assumptions within Ritus formalism.  After a lengthy but standard procedure, the SDE equation reduces to
\be
1=2e^2\sqrt{2eB} \int \frac{d^3\hat q}{(2\pi)^3} \frac{e^{-{\hat q}_\perp
}}{{\hat q}^2} \frac{1}{q_\parallel^2+m_{dyn}^2}\;,
\ee
where $q^2=q_0^2+q_\perp^2$, $q_\perp^2=q_1^2+q_2^2$ and $\hat Q=Q/\sqrt{2eB}$ for $Q=q_0, \ q_1, \ q_2$. Defining $\alpha=e^2/(4\pi)$, after straightforward integration, the above expression reduces to
\be
1=-\frac{\alpha}{m_{dyn}} e^{-\frac{m_{dyn}^2}{2eB}}\left[i\pi {\rm erf}\left(\frac{im_{dyn}}{\sqrt{2eB}}\right)+{\rm E_i}\left(\frac{m_{dyn}^2}{2eB}\right) \right]\;,
\ee
where $-i{\rm erf}(ix)$ is the error function with complex argument and ${\rm E_i}(x)$ is the exponential integral function.
For consistency of the approximation, we require $eB\gg m$, in such a way that $m_{dym}$ obeys the transcendental relation
\be
1=\frac{\alpha}{m_{dyn}} \log{\left|\frac{2eBe^{-\gamma_E}}{m_{dyn}^2}\right|}\;,
\ee
with $\gamma_E\simeq 0.577216$ being the Euler constant. Thus, 
\be
m_{dyn}=2\alpha W\left( \frac{e^{-\frac{\gamma_E}{2}}\sqrt{2eB} }{2\alpha}\right)\;,\label{mdyn}
\ee
where $W(x)$ is the Lambert $W$ function, {\em i.e.}, the inverse of the function $f(w)=we^w$ for any complex number $w$. In Fig.~\ref{fig2} we display $m_{dyn}$ as a function of $\alpha$ and $eB$. It is positive definite. The result in Eq.~(\ref{mdyn}) was derived in Ref.~\cite{Farakos}  

Inserting the non-perturbative propagator into Eq.~(\ref{corriente2}), we obtain  the charge density 
\be
j^0_{\mathcal A}= -\frac{e^2m_{dyn}B}{4\pi^2} \int_{-\infty}^{\infty} \frac{dp_0}{p_0^2 + m^2_{dyn}} = -\frac{e^2B}{4\pi} , 
\ee
in agreement with well known perturbative results~\cite{Khalilov} identifying $m=m_{dyn}$. Notice that  in Eq.~(\ref{corriente2}), the difference $\left[\vert E_{p,+1}(z)\vert^2 - \vert E_{p,-1}(z)\vert^2\right]$ is such that there exists a neat cancellation of the contribution to the charge density between subsequent Landau levels, and only the LLL contribution prevails. 

The fermion condensate, on the other hand, becomes
\be
\langle \bar {\psi} \psi \rangle^{n=0, \sigma=1}_{\mathcal A} =  \frac{ eB }{4 \pi} \;.
\ee
Equivalent expressions for this quantity have been obtained by different methods~\cite{gusynin} after the same identification $m=m_{dyn}$.

\section{Final remarks}
\label{Final}

In this article we have studied the formation of condensates and vacuum electric current densities of the ground state of massless fermions in (2+1)-dimensions by homogeneous and inhomogeneous magnetic fields. These quantities were extracted directly from the fermion propagator. The effects of an external magnetic field  were included within the Ritus eigenfunctions approach~\cite{Ritus}, which was generalized to incorporate magnetic fields of arbitrary spatial profile. The class of field configurations that can be considered within this formalism are those for which Eq.~(\ref{pauli}) can be solved~\cite{murguia}, and similar conclusions are expected in all these cases~\cite{Dunne}. General expressions for $\langle\bar\psi\psi\rangle$ and $j^\mu$ are presented in Eqs.~(\ref{condensado2})-(\ref{corr2}). 
Although we have worked out explicitly the derivation of these quantities only in the irreducible representation of the Dirac matrices, Eq.~(\ref{primera}), ensuring that only one spin orientation for fermions enter in the LLL, we have specified how our findings can be translated to the second inequivalent~(\ref{segunda}) and the reducible~(\ref{reducible}) representations, where we have also considered  parity non-invariant mass terms. 

In the large magnetic flux regime, we have seen that both $\langle\bar\psi\psi\rangle$ and $j^\mu$ are proportional to the external field. In perturbation theory, the local relation between the inhomogeneous condensate and the flux can be interpreted as a local form of the Aharonov-Casher relation~\cite{AC}, as anticipated earlier in~\cite{Dunne}. The induced current density, in turn, has the form
\be
j^\mu(x)=- \frac{e^2}{4\pi} {\rm sgn} (m) ^*F^\mu(x),
\ee
with $^*F^\mu(x)=\epsilon^{\mu\nu\lambda}F_{\nu\lambda}(x)/2$. Therefore, it is gauge invariant and conserved. 
We observe that there exists a LLL dominance for the formation of the condensate and charge density for  intense inhomogeneous magnetic fields. For the non-perturbative uniform condensate and induced current results, an interesting question that naturally arises is whether in the magnetic catalysis scenario in (2+1)-dimensions a Chern-Simons term of non-perturbative origin should be considered to cure the anomaly. However, for uniform fields, such a term vanishes formally for our choice of the vector potential $A_\mu$~\cite{Redlich}. Thus the computation of the induced current alone is not sufficient to deduce the presence of such a term in the complete effective action. The effective action $\Gamma$ for the gauge field has to be computed simultaneously and then, the Chern-Simons term can be inferred from~\cite{Redlich}
\be
\frac{\delta \Gamma}{\delta A_\mu}= j^\mu.
\ee
This and other  additional effects of homogeneous and inhomogeneous magnetic fields, like the dynamical generation of mass and anomalous magnetic moment~\cite{cubanos}, for  the Lagrangian~(\ref{Diraclagredsep}) are currently being considered and will be presented elsewhere.

\section*{Acknowledgments}
We are indebted to Alejandro Ayala, Adnan Bashir, Efra\'{\i}n Ferrer, Vivian de la Incera, Gabriela Murgu\'{\i}a and Angel S\'anchez for valuable discussions and careful reading of the manuscript. AR acknowledges support from SNI and CONACyT grants under project 82230.

\end{document}